\documentclass[twocolumn,showpacs,preprintnumbers,amsmath,amssymb,superscriptaddress]{revtex4}

\usepackage{graphicx}
\usepackage{dcolumn}
\usepackage{bm}

\begin{document}

\title{Antiferromagnetic phase transition in a nonequilibrium lattice of Rydberg atoms}

\author{Tony E. Lee}
\affiliation{Department of Physics, California Institute of Technology, Pasadena, California 91125, USA}
\author{H. H\"{a}ffner}
\affiliation{Department of Physics, University of California, Berkeley, CA 94720, USA}
\author{M. C. Cross}
\affiliation{Department of Physics, California Institute of Technology, Pasadena, California 91125, USA}

\date{\today}

\begin{abstract}
We study a driven-dissipative system of atoms in the presence of laser excitation to a Rydberg state and spontaneous emission. The atoms interact via the blockade effect, whereby an atom in the Rydberg state shifts the Rydberg level of neighboring atoms. We use mean-field theory to study how the Rydberg population varies in space. As the laser frequency changes, there is a continuous transition between the uniform and antiferromagnetic phases. The nonequilibrium nature also leads to a novel oscillatory phase and bistability between the uniform and antiferromagnetic phases.
\end{abstract}

\pacs{}
\maketitle
The behavior of matter far from equilibrium is a fascinating area of study. The presence of driving and dissipation can lead to remarkable phenomena that are not possible in equilibrium. This has motivated much research on nonequilibrium physics in classical systems, such as fluids, chemical reactions, and biological media \cite{cross93,cross09}. An interesting question is: what novel phases appear when a \emph{quantum} system is driven far from equilibrium? Recent cold-atom experiments have studied equilibrium quantum systems in great detail, but they are also a natural setting to study nonequilibrium quantum systems due to the tunability of driving and dissipation \cite{diehl08,diehl10,tomadin11,verstraete09,barreiro11,lee11}. 

In this paper, we study a nonequilibrium many-body quantum system interacting via Rydberg blockade.
A Rydberg atom is one whose electron is excited to a high energy level $n$. The van der Waals interaction between two atoms in identical Rydberg levels scales as $n^{11}$, and this leads to a blockade effect for large $n$: when one atom is excited to the Rydberg state, it prevents nearby atoms from being excited. This is the basis for quantum information processing schemes with Rydberg atoms \cite{jaksch00,lukin01,isenhower10,wilk10,weimer10,saffman10} and a variety of novel phenomena \cite{weimer08,honer10,pupillo10,lesanovsky10,pohl10,schachenmayer10,henkel10,cinti10}. In these schemes, spontaneous emission should be minimized, since it destroys quantum information. On the other hand, spontaneous emission as a source of dissipation may lead to interesting physics, and it can actually be tuned by using different Rydberg levels.

We study a lattice of atoms continuously excited to the Rydberg state and spontaneously decaying back to the ground state. Consider the Rydberg population of each atom, i.e., the fraction of time it spends in the Rydberg state. What is the spatial distribution of the Rydberg population in steady state? Using mean-field theory, we show that as the laser frequency is varied, the system undergoes a continuous transition between a phase with spatially uniform population and a phase with higher population on every other atom. We call the latter the antiferromagnetic phase, since a two-level atom is formally equivalent to a spin-1/2 particle (ground and excited states correspond to down and up spins) \cite{cohen92}. The nonequilibrium nature also leads to a novel oscillatory phase, in which the Rydberg population oscillates periodically in time, and bistability between the uniform and antiferromagnetic phases. Simulations of the full quantum model in 1D (where mean-field theory is least accurate) show that there are short-range antiferromagnetic correlations but not long range order. Our work can be extended to more general dipolar gases and NMR.

First, we describe the Rydberg interaction \cite{saffman10}. Suppose two atoms are in the same Rydberg state $nlj$. There is a dipole-dipole matrix element between $|nljnlj\rangle$ and nearby energy levels, and this interaction shifts the energy of $|nljnlj\rangle$ by an amount $V$. When the atoms are separated by a small distance $R$, the dipolar interaction dominates ($V\approx-C_3/R^3$), but for large distances, the van der Waals interaction dominates ($V\approx-C_6/R^6$). For mathematical convenience, we use the van der Waals interaction and a $|ns_{1/2}ns_{1/2}\rangle$ state, so that the interaction is short range and isotropic. However, it is straightforward to extend the analysis to long-range and anisotropic interactions. The value of $C_6$ depends on $n,l,j$ and is tabulated in Refs.~\cite{singer05,reinhard07,walker08}.

Consider a lattice of atoms that is uniformly excited by a laser from the ground state to a Rydberg state. The atoms are assumed to be fixed in space. Since the van der Waals interaction decreases rapidly with distance, we assume nearest-neighbor interactions. Let $|g\rangle_j$ and $|e\rangle_j$ denote the ground and Rydberg states of atom $j$. The Hamiltonian in the interaction picture and rotating-wave approximation is ($\hbar=1$)
\begin{eqnarray}
H&=&\sum_j H_j + V\sum_{\langle jk\rangle}\,|e\rangle\langle e|_j\otimes |e\rangle\langle e|_k \;,\label{eq:H}\\
H_j&=&-\tilde{\Delta} \,|e\rangle\langle e|_j+\frac{\tilde{\Omega}}{2}(|e\rangle\langle g|_j+|g\rangle\langle e|_j)\;.
\end{eqnarray}
The second term in Eq.~\eqref{eq:H} is the Rydberg interaction, and $H_j$ is the Hamiltonian for a two-level atom interacting with a laser. $\tilde{\Delta}=\omega_\ell-\omega_o$ is the detuning between the laser and transition frequencies. $\tilde{\Omega}$ is the Rabi frequency, which depends on the laser intensity.

The lifetime of the Rydberg state is limited by several processes: spontaneous emission, blackbody radiation, and superradiance \cite{saffman10}. We account for spontaneous emission from the Rydberg level using the linewidth $\gamma$. When a Rydberg atom spontaneously decays, it usually goes directly into the ground state or first to a low-lying state \cite{day08}; the low-lying states are relatively short-lived, so we ignore them. We also ignore blackbody radiation and superradiance, both of which transfer atoms in a Rydberg level to nearby levels. Blackbody radiation can be minimized by working at cryogenic temperatures \cite{beterov09}, and it is not clear if superradiance is important when the interaction $V$ is large \cite{wang07,day08}. Future treatments could account for them by considering several Rydberg levels instead of just one.

Thus, each atom has two possible states, and the system is equivalent to a dissipative spin model. Previous works have added dissipation to other spin models by coupling each spin to a heat bath; in those works, there is global thermal equilibrium, and the spins are described by an effective partition function \cite{werner05,sperstad10}. However, in quantum optics, dissipation from spontaneous emission leads to a nonequilibrium situation, since the coupling to the heat bath is weak and Markovian \cite{cohen92}. The density matrix for the atoms, $\rho$, is described by a master equation that is local in time:
\begin{eqnarray}
\dot{\rho}&=&-i[H,\rho]+\mathcal{L}[\rho] \;,\label{eq:master}\\
\mathcal{L}[\rho]&=&\gamma\sum_j\left(-\frac{1}{2}\{|e\rangle\langle e|_j,\rho\}+|g\rangle\langle e|_j\,\rho\,|e\rangle\langle g|_j\right)\;.
\end{eqnarray}
The nonequilibrium nature is exhibited in the interplay between unitary and dissipative dynamics \cite{diehl10,tomadin11}, and we are interested in the properties of the steady-state solution of Eq.~\eqref{eq:master}.


Due to the complexity of the full quantum problem, we use mean-field theory. For equilibrium spin models, mean-field theory is useful for determining the existence of different phases \cite{ashcroft76}. Its predictions are accurate in high dimensions but not in low dimensions. For the current nonequilibrium case, we use the approach of Refs.~\cite{diehl10,tomadin11}: factorize the density matrix by site, $\rho=\bigotimes_j\rho_j$, and work with the reduced density matrices, $\rho_j=\mbox{Tr}_{\neq j}\rho$. This accounts for on-site quantum fluctuations but not inter-site fluctuations: for atom $j$, the interaction, $|e\rangle\langle e|_j\otimes\sum_k\, |e\rangle\langle e|_k$, is replaced with the mean field, $|e\rangle\langle e|_j\sum_k\, \rho_{k,ee}$. In high dimensions, this is a good approximation, since fluctuations of the neighbors average out.

\begin{figure}
\centering
\includegraphics[clip,width=2.5 in, bb=55 250 720 515]{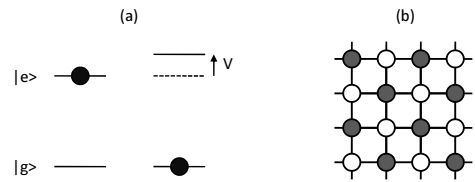}
\caption{\label{fig:lattice} (a) When one atom is excited to the Rydberg state, it shifts the transition frequency of a neighboring atom by $V$. (b) The lattice is divided into two sublattices.}
\end{figure}

Then the evolution of each $\rho_j$ is given by
\begin{eqnarray}
\dot{w}_j&=&-2\tilde{\Omega}\,\mbox{Im }q_j-\gamma(w_j+1) \label{eq:wj}\;,\\
\dot{q}_j&=&i\left[\tilde{\Delta}-\frac{V}{2}\sum_{\langle jk\rangle}(w_k+1)\right]q_j-\frac{\gamma}{2}q_j+i\frac{\tilde{\Omega}}{2}w_j \label{eq:qj}\;,
\end{eqnarray}
where we have defined the inversion $w_j\equiv\rho_{j,ee}-\rho_{j,gg}$ and off-diagonal element $q_j\equiv\rho_{j,eg}$. The Rydberg population $\rho_{j,ee}=(w_j+1)/2$ is the observable measured in the experiment by measuring the scattering rate of each atom. $w_j=-1$ and $1$ mean that the atom is in the ground and Rydberg states, respectively. Equations \eqref{eq:wj} and \eqref{eq:qj} are the Optical Bloch Equations, except that the Rydberg interaction introduces nonlinearity: the detuning for an atom is renormalized by the excitation of its neighbors (Fig.~\ref{fig:lattice}a).

Since the system is dissipative, it will end up at an attracting solution, which can be a fixed point, limit cycle, quasiperiodic orbit, or strange attractor \cite{strogatz94}. (We have not observed the latter two.) We want to know: for given parameter values, how many steady-state solutions are there and are they stable? A solution is stable or unstable if a perturbation to it decays or grows, respectively; the system will end up only in a stable solution. 

Equations \eqref{eq:wj} and \eqref{eq:qj} always have a steady-state solution, in which the Rydberg population is uniform across the lattice ($w_j=w$, $q_j=q$). For some parameter values, this uniform solution is stable, but for others, it is unstable to perturbations of wavelength 2. In the latter case, the lattice divides into two alternating sublattices, and the atoms on one sublattice have higher Rydberg population than the other. Hence an antiferromagnetic pattern emerges from the uniform solution through a dynamical instability. To simplify the discussion here, we keep track of only the two sublattices instead of every site (Fig.~\ref{fig:lattice}b). We stress that the antiferromagnetic transition is not an artifact of using a bipartite lattice, as shown explicitly in the supplemental material \cite{supplement}.


To simplify the equations, we rescale time by $\gamma$ and also rescale the Rabi frequency $\Omega=\tilde{\Omega}/\gamma$, detuning $\Delta=\tilde{\Delta}/\gamma$, and interaction $c=dV/\gamma=-dC_6/\gamma R^6$, where $d$ is the lattice dimension. Labeling the sublattices 1 and 2,
\begin{eqnarray}
\dot{w}_1&=&-2\Omega\,\mbox{Im }q_1-w_1-1\;, \label{eq:w1}\\
\dot{w}_2&=&-2\Omega\,\mbox{Im }q_2-w_2-1\;, \label{eq:w2}\\
\dot{q}_1&=&i\left[\Delta-c(w_2+1)\right]q_1-\frac{q_1}{2}+i\frac{\Omega}{2}w_1 \label{eq:q1}\;, \\
\dot{q}_2&=&i\left[\Delta-c(w_1+1)\right]q_2-\frac{q_2}{2}+i\frac{\Omega}{2}w_2 \label{eq:q2}\;.
\end{eqnarray}
There are six nonlinear differential equations (since $q_1$ and $q_2$ are complex) and three parameters ($\Omega$, $\Delta$, $c$). The uniform version of these equations ($w_1=w_2$, $q_1=q_2$) has been studied before in the context of a medium that interacts with its own electromagnetic field; it is known that there is bistability \cite{hopf84}. We are considering the more general case by letting the sublattices differ.

In the supplemental material \cite{supplement}, we determine the solutions and stabilities for Eqs.~\eqref{eq:w1}-\eqref{eq:q2}. Here, we summarize the main results. Consider first the fixed points, i.e., when $\dot{w}_1=\dot{w}_2=\dot{q}_1=\dot{q}_2=0$. There are two types of fixed points: the uniform fixed points ($w_1=w_2$) correspond to spatially homogeneous Rydberg excitation, while the nonuniform fixed points ($w_1\neq w_2$) correspond to the antiferromagnetic phase, i.e., when one sublattice has higher excitation than the other.

There are either one or three uniform fixed points, corresponding to the real roots of a cubic polynomial,
\begin{eqnarray}
f(w)&=&c^2w^3-c(2\Delta-3c)w^2+\left[\frac{\Omega^2}{2}+\frac{1}{4}\right.\nonumber\\
&&+(\Delta-3c)(\Delta-c)\Big]w+(\Delta-c)^2+\frac{1}{4} \;.\label{eq:f}
\end{eqnarray}
As the parameters change, pairs of uniform fixed points appear and disappear via saddle-node bifurcations. The uniform fixed points never undergo Hopf bifurcations, so we do not expect limit cycles emerging from them \cite{strogatz94}. 

There are up to two nonuniform fixed points, given by the real roots of a quadratic polynomial,
\begin{eqnarray}
g(w)&=&c^2(1+4\Delta^2+2\Omega^2)w^2\nonumber\\
&&-2c[(\Delta-c)(1+4\Delta^2)+(2\Delta-c)\Omega^2]w\nonumber\\
&&+c^2(1+4\Delta^2)-2c\Delta(1+4\Delta^2+2\Omega^2)\nonumber\\
&&+\frac{1}{4}(1+4\Delta^2+2\Omega^2)^2 \;.\label{eq:g}
\end{eqnarray}
The two roots correspond to $w_1$ and $w_2$.
As the parameters change, the two nonuniform fixed points appear and disappear together.

\begin{figure}
\centering
\includegraphics[width=3.5 in]{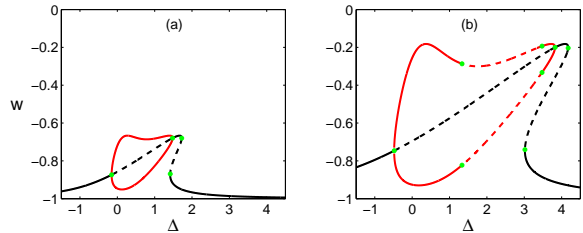}
\caption{\label{fig:bifurcation}Bifurcation diagram showing fixed-point solutions as function of $\Delta$, with $c=5$ and (a) $\Omega=0.5$ and (b) $\Omega=1.5$. The inversion $w$ is -1 (1) when the atom is in the ground (Rydberg) state. Solid (dashed) lines denote stable (unstable) fixed points. Black (red) lines denote uniform (nonuniform) fixed points. Green points denote bifurcations. In (b), the nonuniform fixed points undergo Hopf bifurcations at $\Delta=3.48$ and $1.33$, and there is a stable limit cycle in that interval (shown in Fig.~\ref{fig:soln_phasediagram}a).}
\end{figure}

Since the laser detuning $\Delta$ is the easiest parameter to vary experimentally, we describe what happens as a function of it (Fig. \ref{fig:bifurcation}). Suppose $\Delta$ starts out large and negative. There is one stable uniform fixed point and no other fixed points. As $\Delta$ increases, the uniform fixed point may undergo a pitchfork bifurcation, in which it becomes unstable and the nonuniform fixed points appear. The bifurcation is supercritical, which means that when the nonuniform fixed points appear, they are stable and coincide with the uniform fixed point \cite{strogatz94}. Thus, this is a continuous phase transition between the uniform and antiferromagnetic phases. As $\Delta$ increases further, there is another supercritical pitchfork bifurcation, in which the same uniform fixed point becomes stable again and the nonuniform fixed points disappear. As $\Delta$ increases further towards $\infty$, there is again one stable uniform fixed point and no other fixed points.

\begin{figure}
\centering
\includegraphics[width=3.5 in]{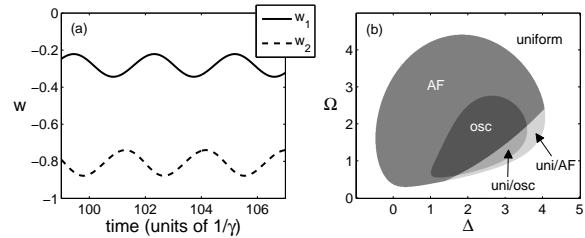}
\caption{\label{fig:soln_phasediagram}(a) Oscillatory steady-state solution (limit cycle) for $c=5$, $\Omega=1.5$, and $\Delta=1.5$. (b) Phase diagram for mean-field theory in $\Omega$,$\Delta$ space for $c=5$. The system is either in the uniform, antiferromagnetic, or oscillatory phase. It can be bistable between uniform and antiferromagnetic phases or between uniform and oscillatory phases.}
\end{figure}

Although the nonuniform fixed points are stable when they appear and disappear, they could become unstable in between via a Hopf bifurcation \cite{strogatz94}. We find numerically that sometimes the nonuniform fixed points do have Hopf bifurcations (Fig.~\ref{fig:bifurcation}b) and give rise to a stable limit cycle, in which $w_1$ and $w_2$ oscillate periodically in time (Fig.~\ref{fig:soln_phasediagram}a). This oscillatory phase is due to the nonequilibrium nature of the system. 


Thus, in mean-field theory, there are three phases: uniform, antiferromagnetic, and oscillatory. Figure \ref{fig:soln_phasediagram}b shows a phase diagram in $\Delta,\Omega$ space. For some parameters, the system is bistable between uniform and antiferromagnetic or between uniform and oscillatory (Fig.~\ref{fig:bifurcation}b); the final state depends on the initial conditions.

\begin{figure}
\centering
\includegraphics[width=3.5 in]{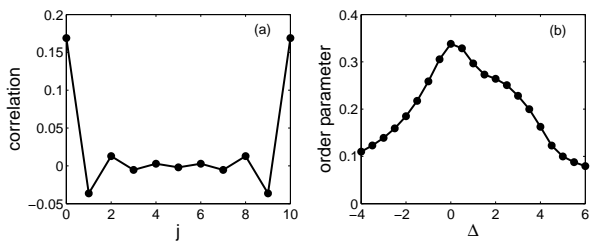}
\caption{\label{fig:simulation} Numerical solution of master equation for 1D chain of length $N=10$ with periodic boundary conditions. Steady state $\rho$ is found after integrating for time $\gamma t=20$. Parameters are $\Omega=1.5$ and $V=5\gamma$, which is equivalent to Fig.~\ref{fig:bifurcation}b. (a) Correlation as a function of distance $j$ for $\Delta=0$. (b) Order parameter as a function of detuning. }
\end{figure}

We also numerically solve the original master equation, Eq.~\eqref{eq:master}, in 1D, where mean-field theory is least accurate. We use fourth-order Runge-Kutta integration to find the steady-state $\rho$ for a chain of length $N=10$. Figure \ref{fig:simulation}a shows the correlation as a function of distance, $\langle E_iE_{i+j}\rangle-\langle E_i\rangle\langle E_{i+j}\rangle$, where $E_i=|e\rangle\langle e|_i$. The rapid decay suggests that there is no long range order in 1D, but the fact that it alternates sign means that there is an antiferromagnetic tendency. We also calculate the order parameter, $[\langle(E_e-E_o)^2\rangle]^{1/2}$, where the operator $E_e=\frac{2}{N}\sum_{i\mbox{\scriptsize{ even}}}E_i$ measures the average Rydberg population on the even sublattice, and $E_o$ does likewise for the odd sublattice. The order parameter measures the difference between the two sublattices: it is 0 when they are identical (uniform phase), but positive when they are different (antiferromagnetic and oscillatory phases). The order parameter is largest for roughly the same parameter space, for which mean-field theory predicts the uniform phase to be unstable (compare Fig.~\ref{fig:simulation}b with Fig.~\ref{fig:bifurcation}b). Thus, mean-field theory captures some qualitative aspects of the full quantum model in 1D, but it remains to be seen whether there is long-range order in higher dimensions, where mean-field theory is more accurate. Also, the prediction of an oscillatory phase suggests that the emitted light has system-wide temporal correlations; indeed, we have found strong photon correlations, which will be reported in detail elsewhere \cite{lee11b}.


Since it is difficult to simulate large systems, experiments with atoms in an optical lattice could provide much information. For example, one can use ${}^{87}\mbox{Rb}$ and a two-photon excitation scheme to go from the ground state $5s_{1/2}$ to the Rydberg state $23s_{1/2}$, which has van der Waals interaction $C_6=-870 \mbox{ kHz } \mu m^6$ \cite{reinhard07} and linewidth $\gamma/2\pi=14.7 \mbox{ kHz}$ at 0 K \cite{beterov09}. A $d$-dimensional lattice with spacing $R=1.5 \mbox{ $\mu$m}$ has interaction strength $V=76$ kHz and $c=5.2d$. The Rydberg population of each atom may be measured by imaging the spontaneously emitted photons; in the antiferromagnetic phase, every other atom fluoresces more. Alternatively, the ground-state population may be measured using repeated projective measurements on a $5s-5p$ transition. A practical setup would be to use a microscope that both produces the lattice and images the atoms \cite{bakr09}.


Thus, a driven-dissipative system of Rydberg atoms has a unique type of antiferromagnetism. The next step is to investigate in more detail how the full quantum model behaves in low dimensions. Our work can be extended to Rydberg states with anisotropic and long-range interactions. Such interactions usually give rise to very rich physics \cite{lahaye09}, so the nonequilibrium version should be interesting. One can also see what happens when the atoms are not fixed on a lattice but free to move; this is reminiscent of classical reaction-diffusion systems \cite{cross93,cross09}. Finally, we note that a system of interacting Rydberg atoms is similar to a system of spins interacting with each other's magnetic dipolar field \cite{jeener99,lin00}. Thus, when an NMR system is made nonequilibrium with continuous driving and spin relaxation, the spins may form a stable pattern in space.

We thank A. Daley and G. Refael for useful discussions. This work was supported by NSF Grant No. DMR-1003337.

\bibliography{rydberg}

\begin{widetext}
\appendix

\begin{center}
{\bf SUPPLEMENTAL MATERIAL}
\end{center}

In this supplement, we provide details on the steady-state solutions of the mean-field model.
In Appendix A, we study the solutions of the bipartite lattice. In Appendix B, we study the solutions of the complete lattice. We use extensively the language of nonlinear dynamics, which is explained in Ref.~\cite{strogatz94app}.

\section{Bipartite lattice} \label{sec:I}
Here, the lattice is bipartite, and we only keep track of the two sublattices, which are labelled 1 and 2. The system is described by
\begin{eqnarray}
\dot{w}_1&=&-2\Omega\,\mbox{Im }q_1-w_1-1\;, \label{eq:app_w1}\\
\dot{w}_2&=&-2\Omega\,\mbox{Im }q_2-w_2-1\;, \label{eq:app_w2}\\
\dot{q}_1&=&i\left[\Delta-c(w_2+1)\right]q_1-\frac{q_1}{2}+i\frac{\Omega}{2}w_1\;, \label{eq:app_q1}\\
\dot{q}_2&=&i\left[\Delta-c(w_1+1)\right]q_2-\frac{q_2}{2}+i\frac{\Omega}{2}w_2\;, \label{eq:app_q2}\;.
\end{eqnarray}
Remember that $w_1,w_2$ are real while $q_1,q_2$ are complex. There are six differential equations and three parameters ($\Omega,\Delta,c$). The equations are symmetric under the transformations $\{w_1,q_1\leftrightarrow w_2,q_2\}$, $\{\Delta,c,q_1,q_2\rightarrow -\Delta,-c,-q^*_1,-q^*_2\}$, and $\{\Omega,q_1,q_2\rightarrow -\Omega,-q_1,-q_2\}$.

We focus on the fixed points of the system, i.e., when $\dot{w}_1=\dot{w}_2=\dot{q}_1=\dot{q}_2=0$. The fixed points are given by the simultaneous roots of two cubic polynomials:
\begin{eqnarray}
f_1(w_1,w_2)&=&(w_1+1)\left\{\frac{1}{4}+[\Delta-c(w_2+1)]^2\right\}+\frac{\Omega^2}{2}w_1 \;,\\
f_2(w_1,w_2)&=&(w_2+1)\left\{\frac{1}{4}+[\Delta-c(w_1+1)]^2\right\}+\frac{\Omega^2}{2}w_2 \;.
\end{eqnarray}
Once $w_1$ and $w_2$ are found, $q_1$ and $q_2$ can be calculated,
\begin{eqnarray}
q_1(w_1,w_2)&=&\frac{-\frac{\Omega}{2}w_1[\Delta-c(w_2+1)]+i\frac{\Omega}{4}w_1}{\frac{1}{4}+[\Delta-c(w_2+1)]^2} \;,\\
q_2(w_1,w_2)&=&\frac{-\frac{\Omega}{2}w_2[\Delta-c(w_1+1)]+i\frac{\Omega}{4}w_2}{\frac{1}{4}+[\Delta-c(w_1+1)]^2} \;.
\end{eqnarray}
Since $f_1(w_1\leq-1,w_2)$ is negative and $f_1(w_1\geq0,w_2)$ is positive and similarly for $f_2$ and $w_2$, we know that the fixed points lie in the range $w_1,w_2\in[-1,0]$.

By combining $f_1$ and $f_2$, we find that the fixed points correspond to the real roots of a fifth-order polynomial $h(w)$, which is too complicated to show here. Fortunately, one can factor it. Note that there are two kinds of fixed points: a uniform fixed point ($w_1=w_2$ and $q_1=q_2$) means that the two sublattices are identical, and a nonuniform fixed point ($w_1\neq w_2$ and $q_1\neq q_2$) means that the system is in the antiferromagnetic phase. The uniform fixed points are given by real roots of a cubic polynomial, $f(w)=f_1(w,w)$,
\begin{eqnarray}
f(w)&=&c^2w^3-c(2\Delta-3c)w^2+\left[\frac{\Omega^2}{2}+\frac{1}{4}+(\Delta-3c)(\Delta-c)\right]w+(\Delta-c)^2+\frac{1}{4} \;.\label{eq:app_f}
\end{eqnarray}
Since the roots of $f$ are also roots of $h$, we know that $f$ is a factor of $h$. Thus, $h(w)=4f(w)g(w)$ and the real roots of the quadratic polynomial $g(w)$ correspond to nonuniform fixed points,
\begin{eqnarray}
g(w)&=&c^2(1+4\Delta^2+2\Omega^2)w^2-2c[(\Delta-c)(1+4\Delta^2)+(2\Delta-c)\Omega^2]w\nonumber\\
&&+c^2(1+4\Delta^2)-2c\Delta(1+4\Delta^2+2\Omega^2)+\frac{1}{4}(1+4\Delta^2+2\Omega^2)^2 \;.\label{eq:app_g}
\end{eqnarray}
Hence, there are at most three uniform fixed points and two nonuniform fixed points. One should think of the two nonuniform fixed points as being a joint pair, since they correspond to $w_1$ and $w_2$.

At this point, the uniform and nonuniform fixed points can be found by numerically solving for the roots of $f$ and $g$, and their stabilities can be determined by calculating the eigenvalues of the Jacobian for each fixed point. However, to obtain general results, we derive as much information as possible analytically without explicitly solving for the fixed points. In particular, we care about the number of each kind of fixed point and their stabilities as a function of the parameters.

\subsection{Number of uniform fixed points} \label{sec:A1}
Here, we examine the number of uniform fixed points, i.e., the number of real roots of $f(w)$. Since $f$ is cubic, it has one or three real roots (two in special cases). Suppose $c$ and $\Delta$ have opposite signs. The polynomial $\tilde{f}(\tilde{w})\equiv f(w=\tilde{w}-1)$ has coefficients with signs $+++-$. By Descartes' rule of signs \cite{anderson98}, $\tilde{f}(\tilde{w})$ has exactly one positive root, which means that $f(w)$ has exactly one root with $w>-1$. Since we know that all roots of $f(w)$ are in the interval $[-1,0]$, this shows that if $c$ and $\Delta$ have opposite signs, there is only one root.

Now we check when $f$ has three roots. Since $f$ cubic, it has three roots when the local maximum and minimum exist and are positive and negative, respectively. Thus, there are three roots if and only if
\begin{eqnarray}
4\Delta^2&>&6\Omega^2+3 \quad\quad \mbox{and}
\end{eqnarray}
\begin{eqnarray}
c&\in&\left(\frac{-2\Delta(18\Omega^2+4\Delta^2+9)-(4\Delta^2-6\Omega^2-3)^\frac{3}{2}}{54\Omega^2},
\frac{-2\Delta(18\Omega^2+4\Delta^2+9)+(4\Delta^2-6\Omega^2-3)^\frac{3}{2}}{54\Omega^2}
\right)\;.
\end{eqnarray}
According to this condition, for large $|\Delta|$, there is exactly one root, i.e., one uniform fixed point, regardless of the sign of $c$.

\subsection{Stability of uniform fixed points} \label{sec:A2}
We check the linear stability of the uniform fixed points to perturbations. Since Eqs.~\eqref{eq:app_w1}-\eqref{eq:app_q2} are symmetric between 1 and 2, the eigenvectors of the Jacobian for uniform fixed points are either symmetric or antisymmetric between 1 and 2. (The symmetric eigenvectors correspond to perturbations that affect 1 and 2 identically, while the antisymmetric eigenvectors represent perturbations that affect 1 and 2 in opposite directions.) This is convenient, because we can check the stability to symmetric and antisymmetric perturbations separately, and the characteristic polynomials are cubic instead of sixth-degree. A uniform fixed point is stable overall if it is stable to both kinds of perturbations.

The Routhe-Hurwitz criterion is very useful here because it can provide stability information without explicitly knowing the fixed point \cite{liu94}. Suppose the characteristic polynomial for a fixed point is cubic: $a_3\lambda^3+a_2\lambda^2+a_1\lambda+a_0$. All the eigenvalues have negative real part if and only if $a_2,a_0,a_1a_2-a_0>0$; this means the fixed point is stable. All the eigenvalues have negative real part except for a pair of purely imaginary roots if and only if $a_2,a_0>0$ and $a_2a_1-a_0=0$; this indicates a Hopf bifurcation \cite{strogatz94app}.

\subsubsection{Stability to symmetric perturbations} \label{sec:A2a}
First we consider the stability of a uniform fixed point to symmetric perturbations. In this case, we can consider a simpler system by letting $w\equiv w_1=w_2$, $q\equiv q_1=q_2$, and
\begin{eqnarray}
\dot{w}&=&-2\Omega\,\mbox{Im }q-w-1\;, \label{eq:app_w}\\
\dot{q}&=&i\left[\Delta-c(w+1)\right]q-\frac{q}{2}+i\frac{\Omega}{2}w \label{eq:app_q}\;,
\end{eqnarray}
whose fixed points are given by the roots of $f(w)$ in Eq.~\eqref{eq:app_f}. Once $w$ is found, $q$ is given by
\begin{eqnarray}
q&=&\frac{-\frac{\Omega}{2}w[\Delta-c(w+1)]+i\frac{\Omega}{4}w}{\frac{1}{4}+[\Delta-c(w+1)]^2}\;.
\end{eqnarray}
The characteristic polynomial for a fixed point $w,q$ is
\begin{eqnarray}
\alpha(\lambda)&=&\lambda^3+2\lambda^2+\left\{[\Delta-c(w+1)]^2+\Omega^2+\frac{5}{4}-2\Omega c \,\mbox{Re }q \right\}\lambda+[\Delta-c(w+1)]^2+\frac{\Omega^2}{2}+\frac{1}{4}-2\Omega c \,\mbox{Re }q \;.\label{eq:app_alpha}
\end{eqnarray}
For $\alpha(\lambda)$, we see that $a_2>0$ and $a_2a_1-a_0=a_0+\Omega^2+2$. According to the Routhe-Hurwitz criterion given above, the fixed point is stable to symmetric perturbations if and only if $a_0>0$. Also, since $a_2a_1-a_0>a_0$, there is never a Hopf bifurcation from symmetric perturbations for any uniform fixed point.

Now suppose $c$ and $\Delta$ have opposite signs. We showed in Sec.~\ref{sec:A1} that in this case, there is one uniform fixed point. We also see that $2\Omega c\,\mbox{Re }q\leq0$ in this case and hence $a_0>0$. Thus the single uniform fixed point is stable to symmetric perturbations.

Since $f$ is cubic, fixed points of Eqs.~\eqref{eq:app_w}-\eqref{eq:app_q} appear and disappear through saddle-node bifurcations as the parameters change \cite{strogatz94app}. In a saddle-node bifurcation, two fixed points with opposite stabilities appear or disappear together. Also, since there is never a Hopf bifurcation, a given fixed point has the same stability as the parameters change.

These statements allow us to deduce the stabilities of all the uniform fixed points. Suppose we start with $c$ and $\Delta$ having opposite signs. There is a single uniform fixed point and it is stable as shown above. As we change the parameters, eventually it collides with another uniform fixed point via a saddle-node bifurcation, so the second fixed point must be unstable. Furthermore, that unstable fixed point undergoes a saddle-node bifurcation with a third fixed point, so the third fixed point must be stable, even when it is the only fixed point. (Remember that these stabilities are with respect to symmetric perturbations.)

To summarize, when there is one uniform fixed point, it is stable to symmetric perturbations. When there are three uniform fixed points, the outer two (highest and lowest values of $w$) are stable to symmetric perturbations, while the inner one (middle value of $w$) is unstable to symmetric perturbations. Of course, the outer fixed points could be unstable to antisymmetric perturbations, which is what we check in the next section. 


\subsubsection{Stability to antisymmetric perturbations}

We now check when the uniform fixed points become unstable to antisymmetric perturbations. Let the fixed point be $w,q$. We consider antisymmetric perturbations around it: $w_1=w+\delta w$, $w_2=w-\delta w$, $q_1=q+\delta q$, and $q_2=q-\delta q$. By plugging into Eqs.~\eqref{eq:app_w1}-\eqref{eq:app_q2} and linearizing for small $\delta w,\delta q$, we find the characteristic polynomial,
\begin{eqnarray}
\beta(\lambda)&=&\lambda^3+2\lambda^2+\left\{[\Delta-c(w+1)]^2+\Omega^2+\frac{5}{4}+2\Omega c \,\mbox{Re }q \right\}\lambda+[\Delta-c(w+1)]^2+\frac{\Omega^2}{2}+\frac{1}{4}+2\Omega c\,\mbox{Re }q \;,\label{eq:app_beta}
\end{eqnarray}
which is similar to Eq.~\eqref{eq:app_alpha}. For $\beta(\lambda)$, we see that $a_2>0$ and $a_2a_1-a_0=a_0+\Omega^2+2$. According to the Routhe-Hurwitz criterion given above, the fixed point is stable to antisymmetric perturbations if and only if $a_0>0$, i.e.,
\begin{eqnarray}
[\Delta-c(w+1)]^2+\frac{\Omega^2}{2}+\frac{1}{4}+2\Omega c\,\mbox{Re }q>0 \;.\label{eq:stabletoantisym}
\end{eqnarray}
We can simplify this using the fact that the fixed point satisfies $f(w)=0$: the uniform fixed point is stable to antisymmetric perturbations if and only if
\begin{eqnarray}
\phi(w)\equiv c^2w^2+2c^2w+c^2-\Delta^2-\frac{\Omega^2}{2}-\frac{1}{4}<0 \;.\label{eq:app_phi}
\end{eqnarray}
So when a uniform fixed point is on the verge of instability, it satisfies
\begin{eqnarray}
w&=&-1+\frac{\sqrt{1+4\Delta^2+2\Omega^2}}{2|c|} \;. \label{eq:app_vergeuniunstable}
\end{eqnarray}
For large $|\Delta|$, $\phi<0$, so the one uniform fixed point that exists is stable to both symmetric and antisymmetric perturbations.

Note that since $a_2a_1-a_0>a_0$, there is never a Hopf bifurcation from antisymmetric perturbations. Since we already ruled out Hopf bifurcations from symmetric perturbations, we conclude that uniform fixed points never have Hopf bifurcations.

\subsection{Nonuniform fixed points}

The nonuniform fixed points are given by the roots of the quadratic polynomial $g(w)$ in Eq.~\eqref{eq:app_g}. The two nonuniform fixed points appear and disappear together as the parameters change. The symmetry in Eqs.~\eqref{eq:app_w1}-\eqref{eq:app_q2} between 1 and 2 means that the two nonuniform fixed points have the same stability. Thus, the nonuniform fixed points must appear via a pitchfork bifurcation from a uniform fixed point as the parameters change \cite{strogatz94app}. In other words, the nonuniform fixed points intersect with a uniform fixed point, which changes stability at the intersection. (In Sec.~\ref{sec:A4}, we will determine whether the bifurcation is supercritical or subcritical and which uniform fixed point is involved.)

The intersection of the nonuniform fixed points with a uniform fixed point can be shown explicitly. From $g(w)$, the nonuniform fixed points exist if and only if $\zeta(\Omega,\Delta,c)<0$, where
\begin{eqnarray}
\zeta(\Omega,\Delta,c)&=&16(1+2\Omega^2)\Delta^4-32c\Omega^2\Delta^3+8(1+2\Omega^2)^2\Delta^2-8c\Omega^2(1+2\Omega^2)\Delta+(1+2\Omega^2)^3-4c^2\Omega^4\;.
\end{eqnarray}
On the verge of the appearance of the nonuniform fixed points, $\zeta=0$ and the root of $g(w)$ is
\begin{eqnarray}
w&=&-1+\frac{\Delta}{c}+\frac{\Omega^2}{1+4\Delta^2+2\Omega^2} \;. \label{eq:app_nonappear}
\end{eqnarray}
One can show that Eqs.~\eqref{eq:app_vergeuniunstable} and \eqref{eq:app_nonappear} are equal using $f(w)=0$. Thus, the change in stability of a uniform fixed point coincides with the appearance of the nonuniform fixed points.

In the case when $|\Delta|$ is large, $\zeta\sim\Delta^4$, so nonuniform fixed points do not exist.

There is a convenient sufficient condition for the existence of the nonuniform fixed points. If $\zeta(\Delta=0)=(1+2\Omega^2)^3-4c^2\Omega^4<0$, there is a range of $\Delta$ around $\Delta=0$, for which the nonuniform fixed points exist. In other words, if
\begin{eqnarray}
|c|&>&\frac{(2\Omega^2+1)^{\frac{3}{2}}}{2\Omega^2}\;,
\end{eqnarray}
there is a range of $\Delta$ around $\Delta=0$, for which nonuniform fixed points exist.

\subsection{Connection between uniform and nonuniform fixed points} \label{sec:A4}
We describe what happens as $\Delta$ changes. Without loss of generality (due to symmetry), assume $c>0$. Suppose $\Delta$ starts out large and negative. There is one uniform fixed point and it is stable. Nonuniform fixed points do not exist. Then let $\Delta$ increase. At some point, a uniform fixed point may undergo a pitchfork bifurcation: it becomes unstable to antisymmetric perturbations and the nonuniform fixed points appear. The fact that the nonuniform fixed points exist when the uniform fixed point is unstable indicates that the bifurcation is supercritical \cite{strogatz94app}. 

When the pitchfork bifurcation happens, there may be three uniform fixed points. Which one undergoes the bifurcation? Since $\phi(w)\sim w^2$ in Eq.~\eqref{eq:app_phi}, as $\Delta$ decreases, the first uniform fixed point to go unstable must be the upper one (the one with the highest $w$). Furthermore, since there can be at most two nonuniform fixed points, only the upper uniform fixed point may be unstable to antisymmetric perturbations. We conclude that when there are three uniform fixed points, only the upper one may undergo a pitchfork bifurcation, and the lower one is stable. According to Sec.~\ref{sec:A2a}, the middle uniform fixed point is always unstable.

It is possible that the pitchfork bifurcation happens when there is only one uniform fixed point. Then obviously that fixed point must undergo the pitchfork bifurcation.

Since the pitchfork bifurcation is supercritical, the nonuniform fixed points are stable when they appear.

Then let $\Delta$ increase towards $\infty$. Eventually $\zeta>0$ and $\phi<0$ again. This indicates that there was another pitchfork bifurcation in which the nonuniform fixed points disappeared and the uniform fixed point became stable again.

Note that we have not determined here whether the nonuniform fixed points ever have Hopf bifurcations. As stated in the main text, we found numerically that they do sometimes have Hopf bifurcations.


\section{Complete lattice}
Here, we study the solutions for the complete lattice, keeping track of every site instead of lumping them into sublattices. We show that there is a dynamical instability, in which the uniform steady state becomes unstable to perturbations of wavelength 2. The solutions of the complete lattice are the same as the solutions using sublattices, so the antiferromagnetic transition is not an artifact of using a bipartite lattice.

The approach is to find the uniform steady state and see when it becomes unstable to perturbations. Consider a $d$-dimensional lattice with $N$ sites in each direction. Let $\vec{n}$ be a $d$-dimensional position vector. The system is described by
\begin{eqnarray}
\dot{w}_{\vec{n}}&=&-2\Omega\,\mbox{Im }q_{\vec{n}}-w_{\vec{n}}-1 \;,\label{eq:app_wn}\\
\dot{q}_{\vec{n}}&=&i\left[\Delta-b\sum_{\langle {\vec{m}}{\vec{n}}\rangle}(w_{\vec{m}}+1)\right]q_{\vec{n}}-\frac{q_{\vec{n}}}{2}+i\frac{\Omega}{2}w_{\vec{n}} \label{eq:app_qn}\;,
\end{eqnarray}
where $b=V/2\gamma=c/2d$ is the nearest-neighbor interaction.

The uniform steady state ($w_{\vec{n}}=w$, $q_{\vec{n}}=q$) is given by the fixed points of the system,
\begin{eqnarray}
\dot{w}&=&-2\Omega\,\mbox{Im }q-w-1 \;,\label{eq:app_wuni}\\
\dot{q}&=&i\left[\Delta-2db(w+1)\right]q-\frac{q}{2}+i\frac{\Omega}{2}w\;.\label{eq:app_quni}
\end{eqnarray}
These equations are the same as Eqs.~\eqref{eq:app_w}-\eqref{eq:app_q} but with $c=2db$. Thus, we can use the results of Sec.~\ref{sec:A2a}. The uniform steady state is given by the real roots of $f(w)$ in Eq.~\eqref{eq:app_f}, and there are one or three solutions. The Jacobian of Eqs.~\eqref{eq:app_wuni}-\eqref{eq:app_quni} determines the stability of a uniform solution to uniform perturbations, i.e., an identical offset to every site.

Now we consider perturbations $\delta w_{\vec{n}},\delta q_{\vec{n}}$ around the uniform steady state:
\begin{eqnarray}
w_{\vec{n}}&=&w+\delta w_{\vec{n}} \;,\\
q_{\vec{n}}&=&q+\delta q_{\vec{n}} \;.
\end{eqnarray}
We write them in terms of Fourier components $\delta \tilde{w}_{\vec{k}},\delta \tilde{q}_{\vec{k}}$:
\begin{eqnarray}
\delta w_{\vec{n}} &=& \frac{1}{N}\sum_{\vec{k}} e^{i\vec{k}\cdot \vec{n}} \delta \tilde{w}_{\vec{k}} \;,\\
\delta q_{\vec{n}} &=& \frac{1}{N}\sum_{\vec{k}} e^{i\vec{k}\cdot \vec{n}} \delta \tilde{q}_{\vec{k}} \;,\\
\vec{k}&=&(k_1,k_2,\ldots,k_d) \;,\\
k_\ell&=&\frac{2\pi}{N}j\;, \quad\quad j=0,\ldots,N-1\;.
\end{eqnarray}
We write Eqs.~\eqref{eq:app_wn}-\eqref{eq:app_qn} in terms of $\delta \tilde{w}_{\vec{k}},\delta \tilde{q}_{\vec{k}}$ and linearize. The Fourier components are uncoupled from each other and each component $\vec{k}$ is described by three differential equations:
\begin{eqnarray}
\delta\dot{\tilde{w}}_{\vec{k}}&=&i\Omega(\delta\tilde{q}_{\vec{k}}-\delta\tilde{q}^*_{-\vec{k}})-\delta\tilde{w}_{\vec{k}} \;,\\
\delta\dot{\tilde{q}}_{\vec{k}}&=&-2ibq\left(\sum_{\ell=1}^d \cos k_\ell\right)\delta\tilde{w}_{\vec{k}} +i\frac{\Omega}{2}\delta\tilde{w}_{\vec{k}} + \left\{i[\Delta-2db(w+1)]-\frac{1}{2}\right\}\delta\tilde{q}_{\vec{k}} \;,\\
\delta\dot{\tilde{q}}^*_{-\vec{k}}&=&2ibq^*\left(\sum_{\ell=1}^d \cos k_\ell\right)\delta\tilde{w}_{\vec{k}} -i\frac{\Omega}{2}\delta\tilde{w}_{\vec{k}} + \left\{-i[\Delta-2db(w+1)]-\frac{1}{2}\right\}\delta\tilde{q}^*_{-\vec{k}}\;.
\end{eqnarray}
The characteristic polynomial for component $\vec{k}$ is
\begin{eqnarray}
\eta(\lambda)&=&\lambda^3+2\lambda^2+\left\{[\Delta-2db(w+1)]^2+\Omega^2+\frac{5}{4}-4\Omega b \,\mbox{Re }q\sum_{\ell=1}^d \cos k_\ell \right\}\lambda \\
&&+[\Delta-2db(w+1)]^2+\frac{\Omega^2}{2}+\frac{1}{4}-4\Omega b \,\mbox{Re }q \sum_{\ell=1}^d \cos k_\ell\;.
\end{eqnarray}
Note the similarity to Eqs.~\eqref{eq:app_alpha} and \eqref{eq:app_beta}. Now we use the Routhe-Hurwitz criterion given in Sec.~\ref{sec:A2}. For $\eta(\lambda)$, $a_2>0$ and $a_2a_1-a_0=a_0+\Omega^2+2$. Thus, the uniform steady state is stable to a perturbation with wave vector $\vec{k}$ if and only if
\begin{eqnarray}
[\Delta-2db(w+1)]^2+\frac{\Omega^2}{2}+\frac{1}{4}-4\Omega b \,\mbox{Re }q \sum_{\ell=1}^d \cos k_\ell>0\;.
\end{eqnarray}

Suppose the uniform steady state satisfies $4\Omega b \,\mbox{Re }q>0$. As the parameters change, the first mode to go unstable is the one with all $k_\ell=0$. This corresponds to uniform perturbations that simply offset the uniform solution. Thus we identify this uniform steady state as the unstable fixed point of Eqs.~\eqref{eq:app_wuni}-\eqref{eq:app_quni}. Based on the discussion in Sec.~\ref{sec:A2a}, this uniform solution is actually always unstable when it exists.

Now suppose the uniform steady state satisfies $4\Omega b \,\mbox{Re }q<0$. As the parameters change, the first mode to go unstable is the one with all $k_\ell=\pi$. This corresponds to perturbations of wavelength 2, i.e., antiferromagnetic perturbations. The uniform steady state is unstable to this mode when
\begin{eqnarray}
[\Delta-2db(w+1)]^2+\frac{\Omega^2}{2}+\frac{1}{4}+4\Omega db \,\mbox{Re }q>0 \;.
\end{eqnarray}
This is the same as Eq.~\eqref{eq:stabletoantisym} with $c=2db$. Thus, the discussion in Sec.~\ref{sec:A4} applies here: as the parameters change, the uniform solution becomes unstable to the antiferromagnetic solution. One can find the antiferromagnetic solution, but that is equivalent to the nonuniform fixed points of Eqs.~\eqref{eq:app_w1}-\eqref{eq:app_q2}, i.e., the real roots of $g(w)$ in Eq.~\eqref{eq:app_g}.

Thus, the complete lattice has the same antiferromagnetic transition as the bipartite lattice.

\end{widetext}

\end{document}